\documentclass{emulateapj}

\shorttitle{Star Formation in Extended Disks}
\shortauthors{Zaritsky \& Christlein}

\begin{document}

\title{On The Extended Knotted Disks of Galaxies}

\author{Dennis Zaritsky\altaffilmark{1,2} and Daniel Christlein\altaffilmark{3,4,5}}
\altaffiltext{1}{Steward Observatory, University of Arizona, Tucson, AZ 85721}
\altaffiltext{2}{Center for Cosmology and Particle Physics, Dept. of Physics, NYU, New York, NY, 10003}
\altaffiltext{3}{Andes Fellow}
\altaffiltext{4}{Departamento de Astronomia, Universidad de Chile, Casilla 36-D, Santiago, Chile}
\altaffiltext{5}{Department of Astronomy, Yale University, P.O. Box 208101, New Haven, CT, 06520}

\email{dzaritsky@as.arizona.edu,christlein@astro.yale.edu}

\begin{abstract}
The stellar disks
of many spiral galaxies are twice as large as generally thought.
We use archival data from the Galaxy Evolution Explorer mission (GALEX) to quantify the
statistical properties of young stellar clusters in the outer, extended disks of a sample
of eleven nearby galaxies. We find an excess of sources between 1.25 and 2 optical radii, $R_{25}$, for 
five of the galaxies, which statistically implies that at least a quarter of such galaxies have
this cluster population (90\% confidence level),
and no significant statistical excess in the sample as a whole beyond
2 optical radii, even though one galaxy (M 83) individually shows such an excess.  Although the
excess is typically 
most pronounced for blue ($FUV -NUV < 1$, $NUV < 25$) sources, there is also an excess
of sources with redder colors. Although from galaxy to galaxy the number of sources
varies significantly, on average, the galaxies with such
sources have $75 \pm 10$ blue sources at radii
between 1.25 and 2$R_{25}$.
In addition, the radial distribution is consistent with the extended dust emission observed in the far IR
and with the properties of $H\alpha$ sources,  assuming a constant cluster formation rate
over the last few hundred Myrs. All of these results suggest that the phenomenon of low-level
star formation well outside the apparent optical edges of disks ($R \sim R_{25}$) 
is common and long-lasting. 
\end{abstract}


\section{Introduction}

Simple questions about galaxies can be difficult to answer. For example, the question
``How large are disk galaxies?"  has many answers. One could cite results
regarding the edges of optical disks \citep{vdks}, of the star formation threshold \citep{kennicutt},
of H I disks \citep{roberts}, or of dark matter halos \citep{zaritsky93}. Even within each of those
specific aspects of spiral galaxies
there are ambiguities. For example, recent results have shown both that
the stellar disks are often not sharply truncated \citep{pohlen02,pohlen06}, that star formation
exists beyond the radius of the sharp decline in star formation rate \citep{ferguson98, thilker05},
and that the main ingredient for star formation, molecular gas, exists at large radii \citep{braine}. 
The discovery of radially extended star formation in at least some spiral galaxies has implications
for a range of topics, including the growth and development of disks, star formation
in low-metallcity, low-density environment, and galactic dynamics.
So far, the presentation of extended star formation, either as H II regions \citep{ferguson98}
or UV knots \citep{thilker05,bianchi05, dePaz05}, has focused on the discovery of the
phenomenon, and as such, has been presented in either individual galaxies or
samples of two or three systems. A statistical study was presented by \cite{boissier}, but while
it focuses on integrated luminosity to reach interesting results regarding the star formation 
threshold and initial mass function, it does not treat the star clusters separately. Is there any systematic behavior in extended disk star cluster
formation that we can identify?
In particular, what is the pervasiveness and radial extent of the phenomenon? We use this study 
to lay some groundwork for our kinematic and chemical 
abundance studies of outer-disk H II regions \citep{christlein1,
christlein2,hf}.

\begin{figure*}
\epsscale{1.0}
\plotone{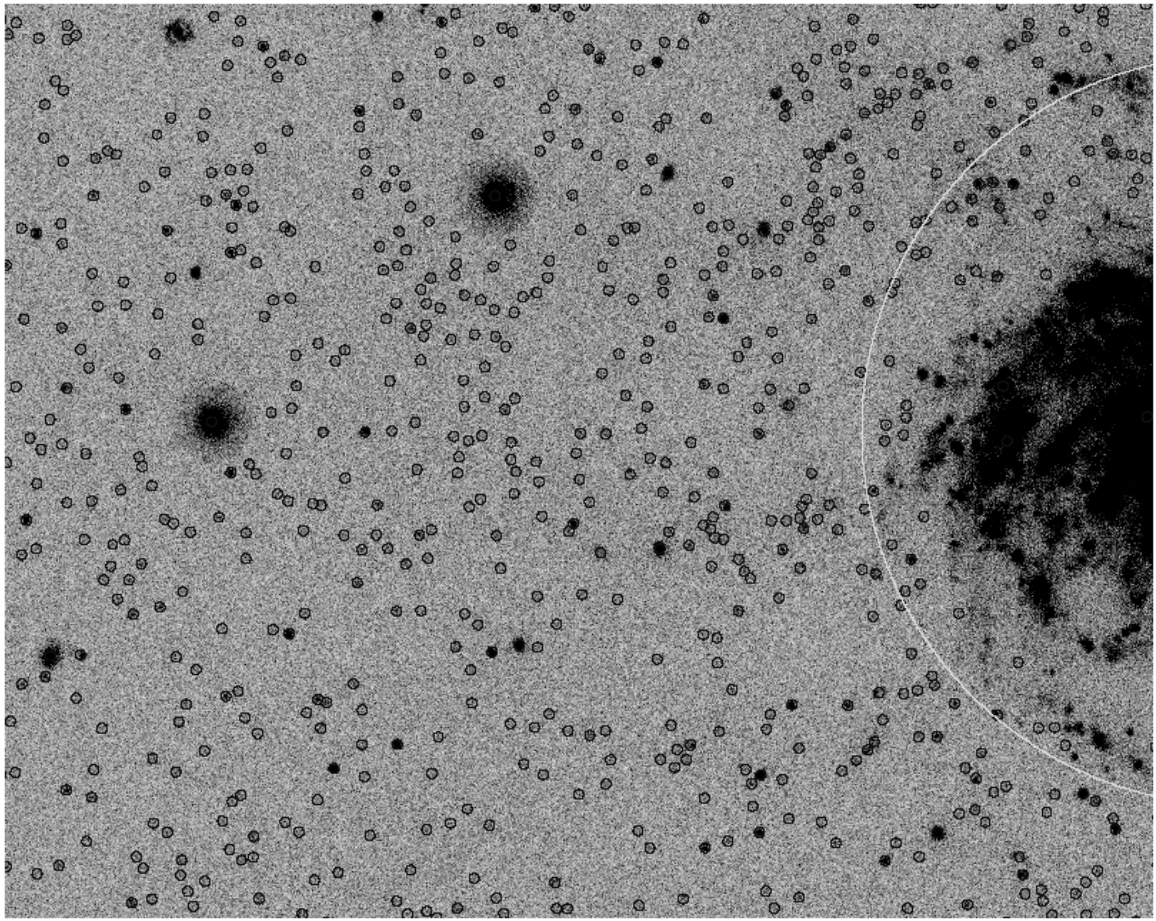}
\caption{
	Source identification in a subfield around NGC 628. Greyscale represents the
	GALEX NUV image (the subsection is 20 arcmin across). Small black circles denote detected sources.
	The large white circle denotes the excluded region in our analysis around NGC 628.
	\label{fig:images}}
\end{figure*}

We use archival data from the GALEX (Galaxy Evolution Explorer) mission to study the phenomenon of UV-selected knots
in the outer regions ($R > R_{25}$) of galaxies. Because our goal is to determine statistical properties valid
for disk galaxies as a class, we will stack the data for a suitable sample of galaxies and examine
the angular cross-correlation function between the UV knots and the parent galaxy. This
study will then provide a measure of the excess number of UV knots in the vicinity
of nearby galaxies and therefore on the prevalence and radial range of the extended star formation. 
In \S2 we describe the data and our treatment of them. In \S3 we describe the cross-correlation
results and in \S4 we discuss some implications.

\section{Sample Selection and GALEX Data}

We select our sample of face-on, well-resolved, disk galaxies 
from the extensive catalog of GALEX galaxy observations
published by \cite{dePaz06}. To select face-on galaxies,
we
define the inclination using the ratio of major, $a$, to minor axis, $b$, appropriate 
for projected circular disks and then select galaxies with 
an inferred inclination less than 45$^\circ$. To select galaxies of large angular extent 
that  still lie well within a single GALEX pointing, so that we obtain a reliable measurement
of the background source density from the same image, we retain only those galaxies with 
$2 < a < 10$ arcmin. The semi-major axis is taken from the compilation of \cite{dePaz06} and
refers to $R_{25}$, which we also interchangeably call the optical radius.
Finally, to select disk galaxies, we require
$0 <$ T-type $\le 7$, where T-Type characterizes morphology as described in
the Third Reference Catalog of Bright Galaxies \citep{rc3} and our criteria span morphological
types Sa to Sd.
These selection criteria result in a sample of 20 galaxies. We then require that 
both high quality far and near UV data exist in the MAST archive at the Space
Telescope Science Institute. Lastly, we reject
NGC 1068 because of the highly non-uniform background in the image that has most likely
variably affected source detection.  
We arrive at the final list of 11 galaxies presented in Table 1 and refer to this as our base
sample.

We refer the reader to \cite{martin05} and \cite{morrissey05} for a description of the mission
and its on-orbit performance.  The images and object catalogs
used here are drawn from the 
database made available by the GALEX team via the MAST archive.
The archival catalogs contain source positions and
photometry in both the $FUV$ (1350 --- 1750 \AA) and $NUV$ (1750 --- 2750 \AA) channels
and were constructed using SExtractor \citep{bertin}. Outside $\sim$ 1.5 optical radii from
the centers of the target galaxies,
the catalogs appear to faithfully recover nearly all visible sources (see Figure \ref{fig:images}). We
discuss this issue in more detail below. The field-of-view is a circle of 1.2$^\circ$ diameter.
The pixel scale is 1.5 arcsec pixel$^{-1}$ and the image point-spread function has a 
FWHM of 4.6 arcsec.
Exposure times range from 541 to 4019 seconds per filter among the
galaxies in our sample. The exposure times are equivalent in the two channels 
for all of our galaxies. Quoted magnitudes are on the AB system and $m=0$ corresponds to 
$1.4 \times 10^{-15}$ and $2.05\times 10^{-16}$ erg s$^{-1}$ cm$^{-2}$ \AA$^{-1}$ for the
$FUV$ and $NUV$ bands, respectively.

We use the $NUV$ astrometry  and aperture photometry to define the object sample and 
use the $FUV$ matched aperture photometry to define the $FUV-NUV$ color. Among the
set of available apertures, we select the 7 pixel aperture as a compromise between including
a significant fraction of the light from the source (7 pixels corresponds to approximately
2 FWHM) and minimizing contamination from nearby sources. These apertures are 
plotted in Figure \ref{fig:images}. We trim the full object
catalog for each galaxy
using a chosen outer image edge, beyond which the catalogs appear to 
often have spurious sources,
and an inside radius from the target galaxy, within which the catalogs
are clearly deficient in detecting sources, perhaps due to source
confusion. The outer cut is defined relative to the field-of-view center and occurs 
typically at a radius of about 1300 pixels. The inner cut is defined relative to the target
galaxy nucleus and is 
typically at a radius of about 1.5 optical radii. The adopted values for each of these
cuts  are based on 
visual inspection of the images and source catalogs and are given in Table 1. An error in
the placement of the inner cut will either cause a spurious drop in counts at small radii (if
the cut is placed at too small a radius), or preclude us from measuring the correlation function
at small radii. An error in 
the outer radius can affect the knot counting statistics (if the choice does not properly
exclude the area with spurious detections), but is clearly identifiable by 
a non-zero level of the background over a range of radii. 

Using the object catalogs, we calculate the angular correlation function, $\xi(r)$, of sources. The
correlation function is defined in the standard manner --- the
excess probability of finding a source at an angular distance $r$ from the reference source.
A random distribution of sources, such as that of the background field,
will average out to a correlation function of zero at all radii (ignoring exotic
effects such as gravitational lensing by the target galaxy 
or obscuration by galactic dust at large radii, for an example 
of the effects of the latter at optical wavelengths see \cite{zaritsky94}). 

\begin{figure}
\epsscale{1.0}
\plotone{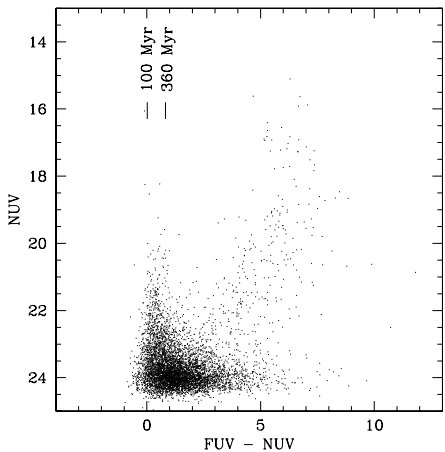}
\caption{
	Color-magnitude diagram of UV knots. The colors of two single stellar population
	models of age 100 and 360 Myr are shown from \cite{bianchi05}.
	\label{fig:cm}}
\end{figure}

\begin{figure}
\epsscale{1.0}
\plotone{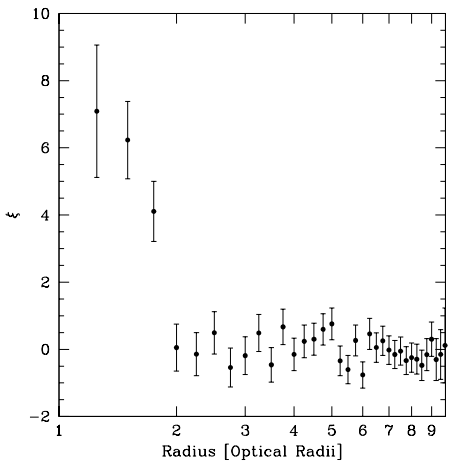}
\caption{
	Correlation function of blue (FUV-NUV $<$ 1) UV-selected knots for the combined final galaxy sample. 
	\label{fig:corr_all}}
\end{figure}

Although a natural choice 
is to use all the objects in the catalog to construct $\xi(r)$, judicious 
choices can enhance the contrast of the extended disk UV knots to the background
population. Previous studies in individual galaxies have stressed the blue UV color
of the most evident population of outer disk sources \citep{thilker05,dePaz05}, so one
advantageous possibility is to focus on the bluer of the detected sources.
Figure \ref{fig:cm} contains the color-magnitude distribution of sources in the NGC 628
field, which is one of the galaxies in the sample for which the extended disk UV knots are
most evident. The Figure includes both blue and red plumes. The blue plume is where the previous
extended disk sources were identified, and we have included labels indicating 
the corresponding color for single stellar populations at two ages
drawn from the modeling of \cite{bianchi05}.
When discussing results below, we will refer to the following populations: ``all" sources
have $FUV-NUV < 10$ and $NUV < 25$, ``bright blue" sources have $FUV -NUV < 1$
and $NUV < 23$, ``blue"  sources have $FUV-NUV < 1$ and $NUV < 25$, ``bright red"
sources have $FUV - NUV > 4$ and $NUV < 23$, and ``faint red" sources have $FUV-NUV > 4$
and $23 < NUV$.

There are various choices that must be made when calculating $\xi(r)$
from a sample of reference sources, as is the case here. First, the radial coordinate
must be defined. The simplest choices are either purely angular (arcsec) or 
physical (kpc), but the galaxies are at different distances and 
of different intrinsic sizes so neither is an optimal choice.
Instead, we use a distance independent scaling radius: optical radii in angular units.
With this choice, $\xi(r)$ now
describes the excess probability of finding a source at a certain number of optical radii from
the central galaxy.
Second, the density of source within different fields, and within a set number of square
optical radii, must be considered.  To account for field-to-field differences, we calculate 
$\xi(r)$ in each field first, and then average the individual galaxy values of $\xi(r)$
to obtain the final $\xi(r)$. This approach has
the apparent disadvantage
that we are not using the entire sample to obtain a highly precise measurement of
the mean background, but instead measure the background from each individual image.
However, it has the advantage that the measurement of $\xi(r)$ is not dominated by
a field with an unusually high background. 
Within each image, we measure the mean background using source counts at radii $>$ 4 optical
radii, which is the farthest that
anyone has yet claimed a detection of UV knots \citep{dePaz05}. 
Our initial size criteria have ensured that we have a sample of galaxies for which a single GALEX
image has significant coverage beyond 4 optical radii from which we can measure 
a background source density.  The final values of $\xi(r)$ are expressed
as a surface density of objects in units of UV knots, or stellar cluster candidates, 
per square optical radius.

\begin{figure}
\epsscale{1.0}
\plotone{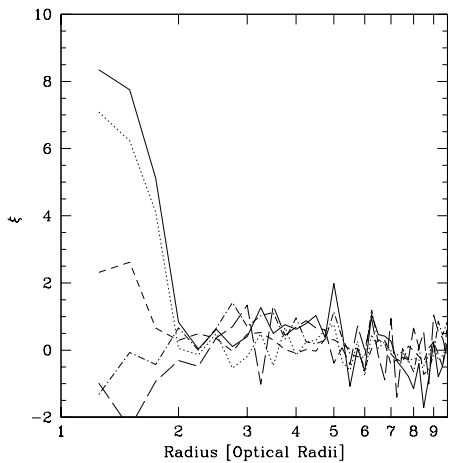}
\caption{
	Correlation function of UV-selected knots as a function of knot selection parameters:
	all sources ($FUV - NUV < 10$, $NUV < 25$) solid line; 
	blue sample ( FUV - NUV $< 1$ and NUV $< 25$) dotted line; bright blue sample 
	(FUV - NUV $<$1 and NUV $< 23$) short dashed line;  bright red sample 
	(FUV - NUV $>$4 and NUV $<$ 23) long dashed line; and faint red sample
	(FUV - NUV $>$ 1 and NUV$>$ 23) dot-dash line.
	\label{fig:corr_color}}
\end{figure}

\section{Results}

In Figure \ref{fig:corr_all} we present $\xi(r)$ for blue sources.
There is a strong signal within two optical radii, and no significant excess of sources thereafter.
The flat correlation function beyond two optical radii
argues against certain potential systematic problems and confirms that defining the mean
background at radii $>$ 4 optical radii is reasonable. The transition in $\xi(r)$ is sharp and
probably underrepresented here due to possible increasing incompleteness in the source catalogs
toward smaller radii (high source densities). Some incompleteness is suggested by the 
negative $\xi(r)$ values at small $r$ for the two red samples of objects (Figure \ref{fig:corr_color}),
which might be expected to be background objects and therefore uniformly distributed. 
However, other explanations for the deficit at small $r$, such as extinction of background sources \citep{zaritsky94},
have not been excluded.

\begin{figure}
\epsscale{1.0}
\plotone{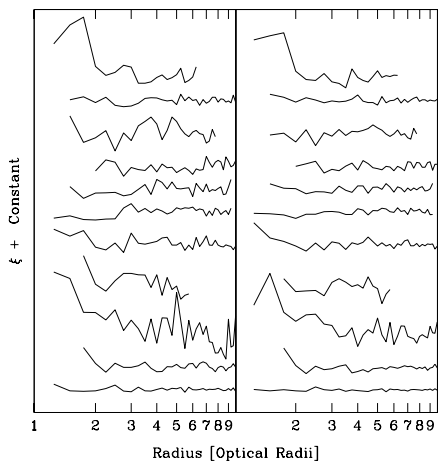}
\caption{
	Correlation function of UV-selected knots for the individual galaxies. Correlation
	functions from all sources in left panel, and only from blue (FUV-NUV $\le 1$) in
	right panel. The correlation functions have been shifted vertically for clarity. The
	galaxies are in the same order as in Table \ref{tab:sample} beginning with NGC 628
	at the top of each panel.
	\label{fig:corr_ind}}
\end{figure}

As previously mentioned, the strength and clarity of the correlation function signal
may depend on the color of the population.
In Figure \ref{fig:corr_color} we plot $\xi(r)$
obtained using all objects and isolating blue, bright blue, bright red, and faint red sources. 
The strongest signal comes from using all the objects, and there is no signal from either
red population. This combination of results suggests that there are some sources with intermediate
colors that contribute signal. However, the stronger signal obtained inside of two optical radii
using all sources is
countered by a poorer baseline for $\xi(r)$ beyond two optical radii due to the poorer
discrimination of background sources. For this reason, we focus on results using the 
blue sample.

\begin{figure}
\epsscale{1.0}
\plotone{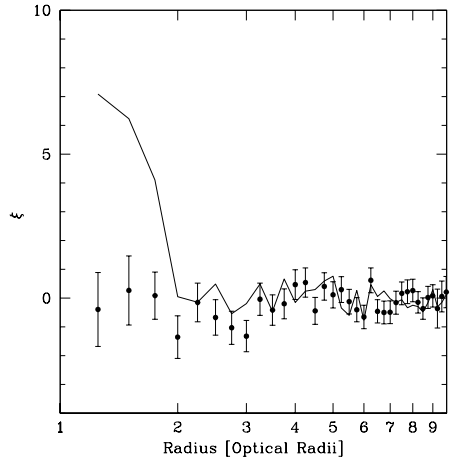}
\caption{Correlation function of UV-selected knots in a selected subsample of galaxies (circles with
error bars). 
We removed the galaxies that show the strongest signature of extended disk sources
(NGC 628, 4303, 4736, 5474, and M 83). The correlation function for the combined sample
is the solid line and shown for comparison.
	\label{fig:corr_low}}
\end{figure}

A key concern in any stacking analysis is whether the final result reflects a universal trend
or is dominated by a small subsample of sources. In Figure \ref{fig:corr_ind} we plot
the correlation functions for each of the 11 galaxies. One of the galaxies, NGC 1672,
has no data at $r < 2$ optical radii and is excluded from further discussion here.
Five galaxies show 
a significant rise in $\xi(r)$ for small $r$ ($r < 2$ optical radii): NGC 628, 
NGC 4303, NGC 4736, M 83, and NGC 5474. We define what is significant here by calculating
the number of  knots detected in the ``all" sample and requiring that this value be sufficiently
large that it is statistically unlikely to happen in one galaxy in a  sample of ten.
In fact, the least significant of these detections (NGC 4736) is 
likely to happen only 4\% of the time by random chance 
(in other words, we have 96\% confidence that our
detections of knots in this galaxies is not due to random fluctuations).
If we remove these five galaxies from the sample and combine
the remainder, we find no significant positive correlation signal within 
two optical radii (Figure \ref{fig:corr_low}).
The five galaxies
for which there is no detected signal are farther on average than those that produce a signal
(17.8 vs. 8.9 Mpc), suggesting
that we may be less sensitive to the knots and more susceptible to source confusion, 
although excluding NGC 7479 from the
calculation of the distances makes the difference less striking (13.5 vs. 8.9 Mpc). 
Using binomial statistics, we calculate the smallest intrinsic fraction of galaxies with knots for which
there is at least a 10\% chance of finding 5 or more positive detections in a sample of 10.
This value is 0.27, so we conclude that with 90\% confidence at least 27\% of spirals
have UV knots at radii between 1.25 and 2 $R_{25}$.

To examine the possibility that we are missing the UV knots in some galaxies due to 
their greater distances in more detail, we consider only blue sources
(those with an age $<$ 360 Myr).
The typical uncertainty in the combined $\xi(r)$ for galaxies that show no evidence individually
of excess sources (Figure \ref{fig:corr_low})
is $\sim 1$ in our units, which suggests 
that even if $\xi(r)$ was a factor of $\sim 3$ lower in these galaxies
we would have significantly detected a difference from zero in at least the innermost radial bin.
The different mean distances of the samples corresponds to a factor of 4 (2.3) in luminosity or 1.5 (0.9) mag, with or (without) NGC 7479.
In Figure \ref{fig:corr_color} we show that even if we raise the magnitude limit by 2 magnitudes
(bright blue vs. blue samples)
we retain a positive signal from the entire sample. We conclude that distance alone is
not responsible for the lack of a detection of extended disk UV-knots in the subsample
of five galaxies.
                                            
Finally, we examine whether we can enlarge the sample by relaxing some of the 
selection criteria. Our criteria, which are superposed on whatever criteria were applied
by the original observers, are angular size, inclination, and morphological type. 
The quantitative cuts we adopt for the first two are in some way arbitrary. For example,
the need for face-on galaxies does not dictate one use an inclination value of 45$^\circ$. 
Morphological type is the only one of the three for which a 
change affects the physical character of the sample, and therefore we do not relax that criteria.

We begin by modestly modifying the inclination cut from 45$^\circ$ to 50$^\circ$. This
change increases our sample size from 11 to 17 (adding NGC 1961, 3351, 3368, 6902, 7741, 
and 7793). Only one of these galaxies individually show a significant rise in the correlation
function at small radii (NGC 6902), where significant is defined as described previously,
and the combined correlation function decreases slightly so that the number of knots for
the entire sample between 1.25 and 2 optical radii declines from 
$33\pm5.8$ to $19.6\pm3.8$. Although the numbers decline, perhaps
because the mean distance of these galaxies is larger than that of the base sample,
the significance of the detection of a source excess remains $> 5 \sigma$.
Increasing the inclination cut by an additional 10$^\circ$ only increases the sample
by 50\%. Large gains in sample size cannot be achieved by modest changes in 
the inclination cut, and the analysis becomes much more ambiguous due to the
projection effects.

We now modify the angular size criteria. The primary consideration here is that angular size
maps onto distance and distance affects our analysis in two ways. First, the knots
of the more distant galaxies are naturally fainter and hence the data do not
reach as far down the luminosity function. This change reduces the number of knots
detected significantly. Second, these fewer knots are embedded in a relatively
more numerous background (because the intrinsically scarce bright knots are now
embedded in the large background population of faint sources). Because selecting
galaxies at larger distances reduces both the absolute numbers of detected knots
and the contrast, we expect a sharp decrease in our ability to measure a positive
correlation function. We have identified a sample of 13 independent galaxies (i.e. 
not in either the base or inclination-extended sample) that have $2 < a < 4$ and
are at distances $<$ 50 Mpc. These were winnowed from a sample of 18 that satisfied
the same criteria. The rejected galaxies either had bright or multiple companions or
there were large scale problems with the images (variable reddening or scattered light). 
At 50 Mpc, the brightest of the knots observed in 
the nearer sample have an apparent magnitude of $\sim$ 24.5. This sample
shows no significant excess (integrated counts from 1.25 to 2 optical radii are
$0.92 \pm 1.65$). To detect this value at 6$\sigma$, comparable to what is done
with the base sample,  would require a sample that is 115 times larger (i.e. 1500 galaxies).
It is therefore  not possible with the GALEX data set to exploit large numbers to offset
the drop in detection sensitivity and relative contrast to the background. We use the results from
the base sample in the discussion.

\section{Discussion}

The identification of extended disks in at least a quarter of disk galaxies indicates
that this is not a fringe phenomenon that affects some systems with
rare specific histories, such as recent interactions (e.g. M 83 and NGC 4625; \cite{chip,bush}).
Instead, we conclude that spiral galaxy disks are commonly at least 2 times larger than indicated
by their optical radii and that in physical size they can extend out to 26 kpc (2 $\times$
average optical radius, which is 12.9 kpc for this sample). While the average optical
radius is 12.9 kpc, the galaxies in the sample radii span from 4.7 to 20.9 kpc. 
The five galaxies for which
we detected correlated signal on an individual basis span the same range, suggesting
that this phenomenon is not limited to systems with abnormally large or small optical radii.

Integrating the correlation function between radii of 1.25 and 2 optical radii results
in a value for the expected excess number of sources of $75 \pm 10$ knots for galaxies with extended 
disks. This number must be treated with some caution because it applies to sources
brighter than $NUV = 25$ for a set of galaxies over a range of distances. 
If between 5 and 10\% of these clusters have associated H$\alpha$ emission, as suggested in preliminary
analysis of observations of knots in NGC 628 \citep{hf}, then a long slit pointing of an edge-on galaxy should yield between three and seven knots. This expectation is in good agreement with the
results of blind slit observations of edge on galaxies by \cite{christlein1}.  This ratio
of H$\alpha$ emitting sources to non-emitting sources is entirely reasonable given the age estimates of
$\sim$ 20 Myr for H II regions and $\sim$ 400 for GALEX knots (see Figure \ref{fig:cm}). 
Establishing the correspondence between the UV-knots and H$\alpha$ sources is
important because we will then be able to connect the kinematics obtained from the H$\alpha$ observations
to the entire population of UV-knots. In particular, \cite{christlein1} have established
that the H$\alpha$ sources follow the disk rotation curve out to $2R_{25}$ and hence
lie in a dynamically stable, differentially rotating disk that continues from the inner disk. 
By inference, the UV knots are also then part of a dynamically cold, extended disk.

\begin{figure}
\epsscale{1.0}
\plotone{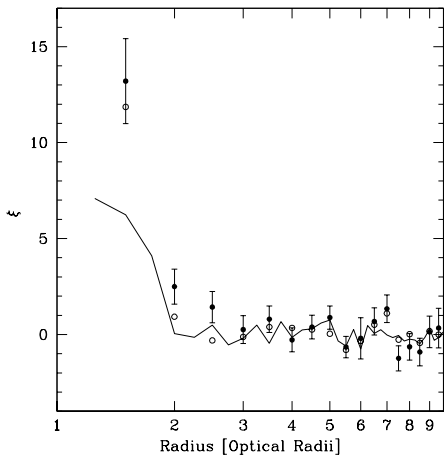}
\caption{Correlation function of UV-selected knots in a selected subsample of galaxies 
that show the strongest signature of extended disk sources
(NGC 628, 4303, 4736, 5474, and M 83). Solid line is the composite for the entire sample
and open circles show the results excluding M 83. 
	\label{fig:corr_high}}
\end{figure}

Connecting the UV-knots to other components, for example H I or dust, will enable 
further study of the evolution of material at these radii. \cite{thilker05} found a general
correspondence between H I filamentary structure 
and UV-knots in M 83, suggesting that H I could be used
to trace the extent of recent star formation in extended disks. We find a correspondence
between the radial range over which we find
UV knots and that over which 100$\mu$m emission was detected by \cite{nelson98}. 
In their most sensitive composite sample, \cite{nelson98} found 100$\mu$m emission with 2$\sigma$
confidence out to 2 optical radii, exactly the radius out to which we find UV knots. Furthermore,
\cite{nelson98} show that the emission they detect comes from a disk-like component, again
consistent with the geometry found here (in combination with the kinematics presented by
\cite{christlein1}). Subsequent studies have found more direct evidence for dust
in  extended disks \citep{popescu} and the correspondence of dust to these areas of star formation
suggests that exotic transport mechanisms from the inner disk are unnecessary.

\begin{figure}
\epsscale{1.0}
\plotone{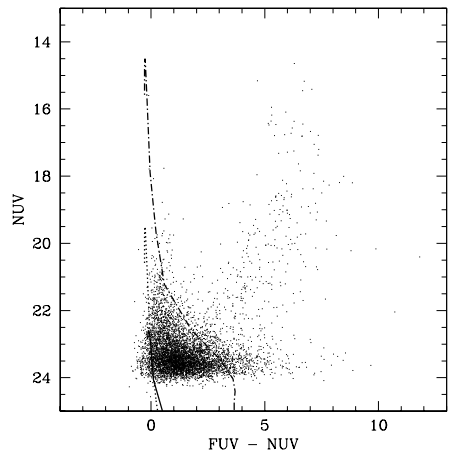}
\caption{Models for the evolution of stellar clusters superposed on properties of
observed UV knots in NGC 628 (corrected to a distance of 8.9 Mpc). 
Solid line represents a clusters with a mass of 1000 M$_\odot$
with a Gaussian burst timescale of 100 Myr, while the dotted line represents the
same mass cluster with a burst timescale of 1 Myr. The dot-dashed line represents
a 10$^6$ M$_\odot$ cluster with a burst timescale of 1 Myr. All models plotted for
an assumed distance equal to the mean of the galaxies that show the excess knot
population (8.9 Mpc). Models do not include any extinction.
	\label{fig:models}}
\end{figure}

To determine what the ages, masses, and contribution to the outer disk surface brightness
of these knots might be,
we have run PEGASE.2 \citep{pegase} models of clusters.
Rather than the standard single stellar population models, we vary the length
of the burst (modeled by a Gaussian star formation rate in time of dispersion, $S$ Myr). We include
models with $S = $ 1 Myr and 100 Myr in Figure \ref{fig:models}. These models are otherwise
standard: Salpeter initial mass function (IMF) for $0.1 < M < 120 M_\odot$, and default PEGASE assumptions
regarding winds, reprocessing of metals, nebular emission, extinction. Changes in the initial
metallicity or IMF do not produce changes that are sufficiently large to impact the 
conclusion discussed here.
We show models of clusters with total masses of 10$^3$ and 10$^6$ $M_\odot$ in Figure \ref{fig:models}. The mass of the 
10$^3$ M$_\odot$ cluster
is only about a factor of two higher than the minimum mass cluster that is likely to have
a star with $M > 20 M_\odot$ (the lowest mass star that will power an H II region), while
the upper mass limit appears to represent an upper bound to the likely cluster 
mass range given the dearth of
observed clusters with NUV $<$ 19. With this range of models, and some reddening
variations, it appears entirely possible to have simulated clusters fill the entirety of the region
in color-magnitude space occupied by the majority
of detected sources.  The two models with different $S$ values demonstrate
the sensitivity of the peak luminosity on the duration of the star formation episode, and 
signal the difficulty in assigning unique ages or masses to candidates.

We conclude from Figure \ref{fig:models} that the sources in the blue plume are likely 
to be clusters younger than 400 Myr in the mass range $10^3 < M < 10^6 M_\odot$, in
concurrence with earlier findings by \cite{thilker05,dePaz05,bianchi05}.
If these clusters form uniformly during those 400 Myr (consistent with the rough $H\alpha$ 
statistics available), and if 50\% of spirals have these populations, then either
50\% of spirals have such sources in their outer disks all the time, these sources
are generated on timescales of a few hundred Myr in spirals, or something in between. 
Either alternative, or a combination, indicates that triggering star formation in the outer
disk is common. If 50\% of spirals have these all the time, and if they typically
have 70 such knots between 1.25 and 2 optical radii, then over the course of 10 Gyr
they will have had 70(10/0.4) = 1750 knots. If these knots have a typical mass of $10^4$
M$_\odot$, then the outer disk at these radii
contains a stellar mass of $\sim$ $1.7 \times 10^7$ M$_\odot$.  
Using the predicted evolution of the V-band magnitude from the models, 
we integrate (using broad, 1Gyr,  time bins
which results in a slight underestimate of the V magnitude) to obtain the V magnitude
of these stars currently. For an assumed distance of 8.9 Mpc we find
V $\sim$ 17.5.
For a middle ground value of $a$ of 4 arcmin, the area between 1.25 and 2 optical radius
equals 27596 sq. arcsec, implying an average surface brightness of $\sim$ 29 mag/sq. arcsec
in V, which is comparable to what the deepest studies reach (although with the
expected surface density gradient expected from 1.25 to 2 optical radii, this population
may be easily detectable at 1.25 optical radii and nearly impossible to detect near 2 optical radii). 
We have ignored reddening due to the low inferred dust content at these radii from
both extinction measurements \citep{zaritsky94} and far infrared luminosity \citep{nelson98}.
We conclude from our modeling that the observed rate of star formation in the outer disk
integrated over a Hubble time does not grossly violate photometric observations of disks, 
and that in some cases my be consistent with the observed optical surface brightness at large radii
\citep{pohlen02,pohlen06}. These models are presented simply to begin a discussion
of the characteristics of this population and clearly await both tighter observational constraints
and more detailed and comprehensive modeling.

\section{Conclusions}

We have studied the distribution of UV-selected sources surrounding a sample of 
nearby, face-on spiral galaxies with the intent of quantifying some properties of the
previously discovered population of star forming regions well beyond the optical
radii of these galaxies \citep{ferguson98,thilker05,dePaz05,bianchi05}. We find the following:

1) Using the cross-correlation function of selected subsamples of UV knots and the 
primary galaxy, we detect an 
excess of sources in 50\% of our sample for 1.25 $< R <$  2 optical radii.
From the integrated counts of knots in this radial range, the detection of knots at extended
radius as a general, although not ubiquitous, phenomenon of spiral galaxies is demonstrated
at $>$ 5$\sigma$ significance. We conclude, with 90\% confidence, that at least
27\% of spirals have such UV source populations.

2) In the combined data set (11 galaxies), we find no evidence for excess sources
beyond 2 optical radii. Individually, only M 83, exhibits an excess of sources beyond
2 optical radii  (in agreement with \cite{thilker05}). Therefore, while such star formation
is relatively common interior to 2 optical radii, it is rare beyond.

3) We find the majority of the correlation signal comes from blue sources ($FUV -NUV < 1$), 
but that there is detectable signal from redder sources ($1 < FUV - NUV )$, indicating
the presence either of some sources older than 400 Myr  or reddened sources. If they
are older than 400 Myr, then they need to be fairly massive ($M > 10^5 M_\odot$). 
Because of the rapid fading of clusters at UV wavelengths, the prominence of
the blue plume does not rule out older clusters, nor suggest the uniqueness
of the current episode of cluster formation.

4) The average number of excess blue sources between 1.25 and 2 optical radii
around spiral galaxies is $33.0\pm5.8$ ($75\pm10$ for those galaxies that show
a statistically significant excess individually).  In combination with the numbers of observed
$H\alpha$ knots (a few in edge on observations \citep{christlein1} and 10\% of UV-selected
sources \citep{hf} in NGC 628), we conclude that the formation rate of these knots could be 
uniform over the last 0.5 Gyr.  

5) Integrating the number of such sources over the lifetime of a galaxy, we obtain an estimate
of the avearge V-band surface brightness of such sources of $\sim 29$ mag/ sq. arcsec. 
Given the expectation
of a gradient in surface density, the surface brightness corresponding to these
sources may be within relatively easy reach at the inner boundary
of the range (1.25 optical radii) and nearly impossible at the outer edge (2 optical radii).
The simple calculation suggests that the extended disks seen in diffuse light in many
galaxies \citep{pohlen02,pohlen06} could arise from a long-term population of evolved
UV-knots.

The outer, or extended, disk environment is well populated by recently ($t < 400$ Myr)
formed stellar clusters in a large fraction of disk galaxies.
As such, it opens several interesting avenues for study, ranging from the physics
of star formation in metal-poor, low density environments (see for example \cite{boissier}),
to their use as kinematic tracers in the disk-halo interface.
These clusters are a common feature of spiral disks and demonstrate that the stellar
disks of galaxies are often twice as large as indicated by the central, high-surface brightness
disk.

\acknowledgments
DZ acknowledges that this research was supported in part by the National
Science Foundation under Grant No. PHY99-07949 during his visit to KITP, a
Guggenheim fellowship,  generous
support from the NYU Physics department and Center for Cosmology and
Particle Physics
during his sabbatical there, NASA LTSA grant 04-0000-0041, NSF
AST-0307482, and
the David and Lucile Packard Foundation.
D.C. gratefully acknowledges financial support from the Fundaci\'on Andes.  
The authors thank the GALEX team for their work on the satellite, instrument, and data
products without which this study would not have been possible, and the MAST team
at STScI for their work in making such data easily accessible to all.


\begin{deluxetable*}{lrrrrrrrrr}
\tabletypesize{\scriptsize}
\tablecaption{Disk Galaxy Sample}

\label{tab:sample}
\tablewidth{0pt}
\tablehead{
\colhead{Name} &
\colhead{$2\times a$}   &
\colhead{$2 \times b$}   &
\colhead{$i$}   &
\colhead{$D$}   &
\colhead{$x_0$}   &
\colhead{$y_0$} &
\colhead{$R_{1}$}&
\colhead{$R_{2}$}
  \\
&
\colhead{[arcmin]}   &
\colhead{[arcmin]}   &
\colhead{[deg]}   &
\colhead{[Mpc]}   &
\colhead{[pix]} &
\colhead{[pix]}  &
\colhead{[a]} &
\colhead{[pix]} \\
}

\startdata
NGC 628&10.5&9.5&25.2&11&1933&1910&1.25&1300\\
NGC 1042&4.7&3.6&40.0&18&2336&1483&1.50&1300\\
NGC 1566&8.3&6.6&37.3&17&1933&1910&1.50&1300\\
NGC 1672&6.6&5.5&33.5&15&2001&1352&2.00&1300\\
NGC 3344&7.1&6.5&23.7&6.9&1930&1908&1.50&1300\\
NGC 3486&7.1&5.2&42.9&12&1932&1910&1.25&1300\\
NGC 4303&6.5&5.8&26.8&17&1932&1909&1.25&1300\\
NGC 4736&11.2&9.1&35.6&5.2&1932&1910&1.65&1300\\
M 83&12.9&11.5&26.9&4.5&935&2456&1.25&1400\\
NGC 5474&4.8&4.3&26.3&6.8&1938&1900&1.75&1300\\
NGC 7479&4.1&3.1&40.8&35&1932&1904&1.25&1300\\
\enddata
\label{tab:sample}
\end{deluxetable*}

\end{document}